# GPU-based ultra fast dose calculation using a finite pencil beam model

**Xuejun Gu[1], Dongju Choi[2], Chunhua Men[1], Hubert Pan[1], Amitava Majumdar[2], and Steve B. Jiang[1]**

[1]Department of Radiation Oncology, University of California San Diego, La Jolla, CA 92037-0843
[2]San Diego Supercomputer Center, University of California San Diego, La Jolla, CA 92093, USA

E-mail: sbjiang@ucsd.edu

Online adaptive radiation therapy (ART) is an attractive concept that promises the ability to deliver an optimal treatment in response to the inter-fraction variability in patient anatomy. However, it has yet to be realized due to technical limitations. Fast dose deposit coefficient calculation is a critical component of the online planning process that is required for plan optimization of intensity modulated radiation therapy (IMRT). Computer graphics processing units (GPUs) are well-suited to provide the requisite fast performance for the data-parallel nature of dose calculation. In this work, we develop a dose calculation engine based on a finite-size pencil beam (FSPB) algorithm and a GPU parallel computing framework. The developed framework can accommodate any FSPB model. We test our implementation on a case of a water phantom and a case of a prostate cancer patient with varying beamlet and voxel sizes. All testing scenarios achieved speedup ranging from 200~400 times when using a NVIDIA Tesla C1060 card in comparison with a 2.27GHz Intel Xeon CPU. The computational time for calculating dose deposition coefficients for a 9-field prostate IMRT plan with this new framework is less than 1 second. This indicates that the GPU-based FSPB algorithm is well-suited for online re-planning for adaptive radiotherapy.

## 1. Introduction

Intensity-modulated radiation therapy (IMRT) is capable of delivering a highly conformal radiation dose to a complex static target volume. However, due to inter-fraction variation of patient anatomy, an optimal IMRT plan designed before the treatment may become less optimal or even totally unacceptable at some point during the treatment course (Yan *et al.*, 2005). With the use of on-board volumetric imaging techniques, this variation can be readily measured before each treatment fraction and then utilized to make useful modification of the original treatment plan (Yan *et al.*, 1997; Jaffray *et al.*, 2002). This procedure is often called adaptive radiation therapy (ART) (Yan *et al.*, 1997; Wu *et al.*, 2002; Birkner *et al.*, 2003; Wu *et al.*, 2004; Mohan *et al.*, 2005; de la Zerda *et al.*, 2007; Lu *et al.*, 2008; Wu *et al.*, 2008; Fu *et al.*, 2009; Godley *et al.*, 2009). One way to implement ART is based on offline re-planning during the treatment course. The schedule for imaging and re-planning is designed based on the modeling of the systematic and random components of the inter-fraction variations. An alternative implementation of ART is based on online re-planning, where a new plan is developed immediately after acquiring the patient's anatomy while the patient is still lying on the treatment couch. The online approach is preferable, as it allows real-time adaptations of the treatment plan to daily anatomical variations. However, it is challenging to implement online ART in clinical practice due to various technical barriers. One such barrier is the requirement of a very fast treatment planning process that can be completed within a few minutes. This constraint is extremely difficult, if not impossible, to satisfy with the central processing unit (CPU)-based computational framework that is currently used in clinical settings.

A typical online adaptive re-planning process includes the following main computational tasks: 1) the reconstruction of in-room computed tomography (CT) images, 2) the segmentation of the target and organs at risk (OARs) in CT images, 3) the computation of the dose distribution on the patient's new geometry, and 4) the optimization of a new treatment plan. Each of these tasks takes a relatively long time under the current CPU-based treatment planning framework and therefore makes online re-planning impractical.

Traditionally, high-speed computation overwhelmingly relies on advance of the processing speed of the CPU. However, the uncertainty of the continuation of exponential growth in processing power as described by Moore's law (Dubash, 2005) has led to renewed interest in multi-core and many-core computational architectures. CPU-clustered traditional supercomputers have been used for solving computationally intensive problems for decades. Despite their great computational power, traditional supercomputers are neither readily available nor accessible to most clinical users due to the prohibitively high cost of facility deployment and maintenance.

The concept of using stream processors in general-purpose graphic processing units (GPUs) depicts an innovative scenario of handling massive floating-point computation and making high-performance computation affordable to general users. GPUs are especially well-suited for problems that can be expressed as data-parallel computations, such as programs of high arithmetic intensity (the ratio of arithmetic operations to memory operations) which need to be executed on multiple data elements in parallel (NVIDIA, 2009). With affordable graphic cards such as NVIDIA's GeForce, GTX, and Tesla series, GPU-based computing has recently been utilized to speed up computational tasks relevant to radiotherapy, such as CBCT reconstruction, deformable image registration, and dose calculation (Sharp *et al.*, 2007; Meihua *et al.*, 2007; Preis *et al.*, 2009; Riabkov *et al.*, 2008; Samant *et al.*, 2008; Xu and Mueller, 2007; Hissoiny *et al.*, 2009; Noe *et al.*, 2008; Jacques *et al.*, 2008). Among them, Jacques *et al* (2008) and Hissoiny *et al* (2009) have explored GPUs for fast dose computation. These two groups focused on the acceleration of the superposition/convolution algorithm. Jacques *et al* (2008) used a combination of a Digital Mars' D program and the Compute Unified Device Architecture (CUDA) development environment. Hissoiny *et al* (2009) modified part of a public domain treatment planning system (PlanUNC) using CUDA and achieved 10-15 times speedup. A larger speedup factor is expected for dose calculation implementations that are designed specifically for GPUs.

By exploring the massive computational resource of GPUs, we are now offered the prospect of overcoming the computational bottleneck of real-time online re-planning. At the University of California San Diego (UCSD), we are developing a supercomputing online re-planning environment (SCORE) based on reasonably priced and readily available GPUs. As part of this effort, we report in this paper the development of a CUDA-based parallel computing framework for ultra fast dose calculation using a finite size pencil beam (FSPB) model. The remainder of this paper is organized as follows. In Section 2.1, the general FSPB model will be introduced and a specific FSPB kernel is described. The CUDA implementation of a general FSPB framework is detailed in Section 2.2. Section 3 presents experimental results of dose calculation in a case of a water phantom and a case of a prostate cancer patient. The computational time is compared between GPU and CPU implementations. Finally, conclusions and discussion are provided in Section 4.

## 2. Methods and Materials

*2.1 FSPB model*

In the process of IMRT inverse planning, a broad beam is divided into small rectangular or square beamlets, and the contribution of each beamlet to every relevant voxel (often called dose deposition coefficient or $D_{ij}$) is calculated using a dose engine. FSPB models are particularly well-suited for such calculations since a beamlet is naturally a finite-size pencil beam, and the accuracy of the models is sufficient in most clinical situations (Bourland and Chaney, 1992; Ostapiak *et al.*, 1997; Jiang, 1998; Jelen *et al.*, 2005). The major assumptions in an FSPB model include: 1) the broad beam from a point source can be divided into identical beamlets; and 2) the dose to a point is the integration of the contribution dose from all beamlets to that point. Mathematically, the dose deposited at a point $P(x, y, z)$ can be written as:

$$D(x,y,z) = \sum_i w_i D_i^p(x,y,z), \qquad (1)$$

where, $D_i^p(x,y,z)$ is the dose distribution of the $i$-th beamlet and $w_i$ is its weighting factor. $D_i^p(x,y,z)$ is also referred to as the dose deposition coefficient or the FSPB kernel. According to the methods proposed by Jiang (1998) and Jelen et al (2005) a general FSPB dose kernel can be formulated as:

$$D^p(x,y,z) = A(\theta,d) F(x',d,z',C), \qquad (2)$$

where, $\theta$ is defined as the angle between the central axis of a beamlet and the central axis of the broad beam, and $d$ is the radiological depth. The patient coordinate system and pencil beam coordinate system used in this work are defined in Figures 1a and 1b, respectively. For the beamlet central axis $\overline{SAB}$, $l$ is the length of $AO'$ and $d$ is the corresponding radiological depth. $A(\theta,d)$ is a factor that accounts for all off-axis effects. $F(x',d,z',C)$ is the beamlet profile at depth $d$ with off-axis effects removed. $x'$ and $z'$ are the coordinates in the pencil beam system, defined on the plane perpendicular to the beamlet central axis. $C$ is a set of parameters defining the shape of the beamlet profile (Jiang, 1998; Jelen *et al.*, 2005).

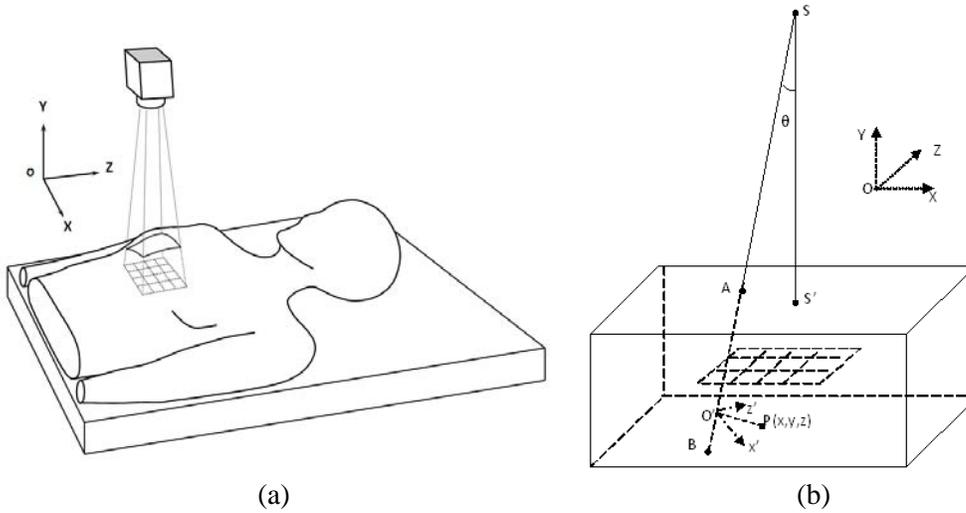

(a)                                                                  (b)

**Figure 1**. (a) An illustration of the FSPB model and the patient-coordinate system. (b) An illustration of the pencil beam coordinate system for calculating the dose from a beamlet $\overline{SAB}$ to point $P$. We define the smallest rectangular prism that encases the patient CT images. $\theta$ is the angle between the central axes of a broad beam ($SS'$) and a beamlet $\overline{SAB}$. $A$ is the entrance point and $B$ is the exit point of the beamlet central axis passing through the rectangular prism. $O'$ is the perpendicular projection of point $P$ onto $\overline{SAB}$. $x'$ and $z'$ are projection of $x$ and $z$ on the plane perpendicular to $\overline{SAB}$.

Under the general FSPB framework defined in Equation (2), there is flexibility in the selection of a beamlet profile function $F(x',d,z',C)$. The solitary requirement is that this selected function satisfies the self-consistency and normalization conditions. Jiang (1998) derived the FSPB dose kernel as the summation of three error functions. Jelen *et al* (2005) implemented an FSPB model by combining several exponential functions. Lin *et al* (2006) constructed a simple and finite-term analytic function by using Boltzmann function. The parameters such as *C* and $A(\theta, d)$ in Equation (2) can be dissolved by fitting depth-dose curves and dose profiles at various depths of broad beams. In this study, we use an error function based FSPB kernel to illustrate our GPU implementation (Jiang, 1998):

$$F(x',d,z',C) = \sum_{i=1}^{3} \frac{c_1^{(i)}(d)}{4} \left[ erf\left(\frac{c_2^{(i)}(d)a/2 - x'}{\sqrt{2}c_3^{(i)}(d)}\right) + erf\left(\frac{c_2^{(i)}(d)a/2 + x'}{\sqrt{2}c_3^{(i)}(d)}\right) \right] \cdot \left[ erf\left(\frac{c_2^{(i)}(d)b/2 - z'}{\sqrt{2}c_3^{(i)}(d)}\right) + erf\left(\frac{c_2^{(i)}(d)b/2 + z'}{\sqrt{2}c_3^{(i)}(d)}\right) \right], \quad (3)$$

where, *a* and *b* are the side lengths of a beamlet's rectangular cross section; $C = \{c_1^{(i)}, c_2^{(i)}, c_3^{(i)}; with\ i = 1, 2, 3\}$ is a set of parameters obtained by fitting broad beam profiles (Jiang, 1998). Note that our implementation allows for this kernel to be easily replaced by any other FSPB kernel.

The dose contribution of a beamlet to a voxel that is far from the beamlet is expected to be very small and can be neglected. To improve computational efficiency, we only compute dose deposition coefficients for voxels inside volumes of interest (VOI). The VOI of a beamlet used in this study is a larger co-central axis beamlet with cross-sectional side length 6 times that of the original beamlet (Fox *et al.*, 2006).

*2.2 CUDA implementation*

Recently, general-purpose computation on GPU (GPGPU) has been greatly facilitated by the development of graphic card language platforms. One example is the CUDA platform developed by NVIDIA (NVIDIA, 2009), which allows the use of an extended C language to program the GPU. In our work, we use CUDA with NVIDIA GPU cards as our implementation platform.

The GPGPU strategy is well-suited for carrying out data-parallel computational tasks, where one program is divided and executed on a number of processor units in parallel. The realization of this single-instruction-multiple-data mode (SIMD) relies on a large number of processing units. General graphic cards such as the GeForce series and the GTX series typically have 32-240 scalar processor units, and the available memory varies from 256MB to 1GB. Recently, NVIDIA introduced special graphic computing processors like the Tesla C1060 that are designed solely for scientific computation. The available memory on the Tesla C1060 is extended to 4 GB. Our implementation is evaluated using a Tesla C1060 card since we believe it has the optimal trade-off between performance and cost in terms of online re-planning applications.

A GPU has to be used in conjunction with a CPU. The CPU serves as the *host* while the GPU is called the *device*. The CUDA platform extends the concept of C functions to *kernels*. A kernel, invoked from the CPU, can be executed *N* times in parallel on the GPU by *N* different CUDA *threads*. For convenience, CUDA threads are grouped to form thread *blocks*, and blocks are grouped to comprise *grids*. The number of blocks and threads has to be explicitly defined when executing a kernel. The threads of a block are grouped into *warps* (32 threads per warp). A warp executes a single instruction at a time across all its threads. On the CUDA platform, the main code runs on the host (CPU), calling kernels that are executed on a physically separate device (GPU). Due to the physical separation of the device and the host, communication between the two cannot be avoided and has to be carefully addressed in CUDA programming.

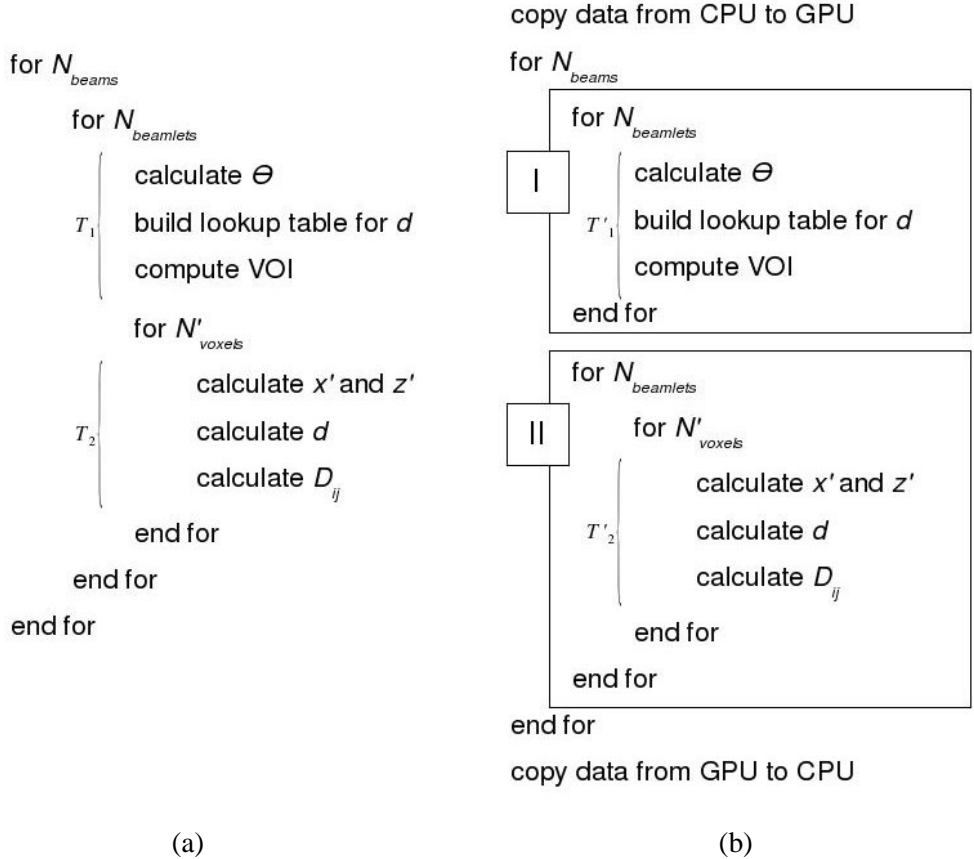

(a)                                                       (b)

**Figure 2**. The pseudocode description of our FSPB algorithm implemented on (a) CPU and (b) GPU. Here the GPU code is parallelized in two different parts. In part I, it is parallelized for all beamlets with both threads and blocks. In part II, the voxels are parallelized with threads and the beamlets are paralleled with blocks. $N'_{voxel}$ refers to the number of voxels contained in a beamlet's VOI.

Figure 2 shows the pseudocode for the implementation of our FSPB model on both CPU and GPU platforms. The CPU implementation (Figure 2a) describes two loops for dose deposition coefficient calculation. The first is over all beamlets, while the second is over all voxels in VOIs. With this scheme, the computational time for a single beam on a single-core CPU can be estimated as:

$$T_{CPU} = N_{beamlets} \cdot (T_1 + N'_{voxels} \cdot T_2), \qquad (4)$$

where, $T_1$ is the computational time over a beamlet loop, and $T_2$ is the time over a loop of a voxel hit by a beamlet. In the GPU pseudocode (Figure 2b), the two corresponding nested loops from the CPU version are separated into two independent ones (Part I and Part II) to efficiently use all GPU multiprocessors. The computational time will be reduced to:

$$T_{GPU} = \frac{N_{beamlets}}{N_{threads}^{(1)} \cdot N_{blocks}^{(1)}} \cdot T_1' + \frac{N_{beamlets}}{N_{blocks}^{(2)}} \cdot \frac{N_{voxels}'}{N_{threads}^{(2)}} \cdot T_2', \quad (5)$$

where, $T_1'$ and $T_2'$ are the computational times when executing on a GPU with a single thread. $T_1'$ and $T_2'$ are numerically equal to $T_1$ and $T_2$ if the GPU and the CPU have the same clock speed. $N_{threads}^{(1)}$ and $N_{threads}^{(2)}$ are the number of threads per block for Part I and Part II, respectively. $N_{blocks}^{(1)}$ and $N_{blocks}^{(2)}$ are the number of blocks for Part I and Part II, respectively. We need to point out that Equation (5) represents the ideal situation. In reality, a GPU has limited numbers of processors and multiprocessors. When the number of threads per block is larger than the warp size (*i.e.*, 32) or the number of blocks is larger than the number of multiprocessors (*i.e.*, 30 for Tesla C1060 GPU), the linear scalability shown in Equation (5) will be broken; GPU computation time will not decreases linearly with the increase of the number of threads per block or the number of blocks.

As shown in Figure 2b, besides GPU computation, there are two additional steps, copying data arrays from the host (CPU) to the device (GPU) prior to GPU computation and copying memory arrays from the device back to the host after computation. Since the data transfer bandwidth within the device (~73,500 MB/s on Tesla C1060) is much higher than that between the device and the host (~2,200 MB/s on PCI Express device Gen 1), frequent communication between the device and the host can significantly impair the overall efficiency. We thus minimize host-device communication by allocating most of the working variables directly in the GPU memory except those carrying input and output data.

In a typical computation operation, the efficiency of code is largely determined by the efficiency of memory management. On a GPU, available memory is divided into constant memory, global memory, shared memory, and texture memory. The constant memory is cached, which requires only one memory instruction (4 clock cycles) to access. The global memory is not cached, requiring 400-600 clock cycles of memory latency per access. However, the available constant memory is limited to 64MB on a general GPU card. Due to this limitation, we store only those arrays with constant values, such as the sources' positions, in the constant memory. Optimal usage of global memory requires coalesced memory access, but the radiological depth calculation cannot avoid random data access. To achieve optimized performance, we use the texture fetch feature of CUDA to access data stored in texture memory.

Conceptually, radiological depth is the integral over the passing length of a beamlet central axis, which can be computed using a ray-tracing algorithm such as Siddon's algorithm (Siddon, 1985). However, doing ray-tracing for each voxel's perpendicular projection $O'$ onto the central axis of each beamlet (Figure 1b) is highly time-consuming.

Thus, we first use a trilinear interpolation method to build a radiological depth lookup table. As shown in Figure 1b, the beamlet central axis enters the smallest encasing rectangular prism at point $A$ and exits at point $B$. We place $n$ equally spaced points on the segment $AB$ and compute the radiological depth from the entrance point $A$ to each of the $n$ points. We can then build a radiological depth lookup table, shown as "build up a lookup table for $d$" in Figure 2a. The radiological depth at perpendicular projection point $O'$ can be obtained with a linear interpolation of two neighboring values stored in the table. Massive memory access occurs when using trilinear interpolation to build the radiological depth lookup table. To avoid this, we place the original CT data in texture memory, which can then be accessed as cached data. The texture fetching possesses a relatively higher bandwidth than global memory access when the coalescing memory accessing pattern cannot be followed. Another issue worth mentioning here is the hardware implementation of the linear interpolation. CUDA offers linear, bilinear, and trilinear interpolation functions at the hardware level. However, due to their insufficient precision, we implemented all interpolation functions in software rather than use the provided hardware functions.

## 3. Experimental Results

*3.1 Water phantom*

The performance of our FSPB implementation on both GPU and CPU platforms was first evaluated with a water phantom experiment, where five co-planar and equally spaced 6 MV beams of $10 \times 10$ cm$^2$ field size were used to irradiate a $30 \times 30 \times 30$ cm$^3$ water phantom. We used the FSPB model of a 6MV beam built and validated by Jiang (1998). The performance of our CUDA code was systemically evaluated by independently varying the size of beamlets and voxels. The CPU code was executed on a 2.27GHz Intel Xeon processor, and the GPU code was run on a NVIDIA Telsa C1060 card. All the testing scenarios are listed in Table 1. In each case, the number of voxels involved in computation (*i.e.*, in all VOIs) was recorded. In Table 1, $T_{CPU}$ is the sequential execution time with the CPU implementation. $T_{GPU}^*$ and $T_{GPU}$ are the execution times on the parallelized GPU implementation accounting for and excluding CPU-GPU data transfer, respectively. The time for data transfer includes copying data from CPU to GPU before GPU computation and from GPU to CPU after GPU computation. This time can be ignored when estimating the required computational time for online re-planning because the data transfer between $D_{ij}$ calculation, automated target/OAR contouring, and plan optimization can occur within the GPU memory. However, when using our GPU implementation for a stand-alone dose calculation, this interval (or at least part of it) should be included in efficiency assessment.

The parallel computation was conducted with CUDA kernels of fixed size. For the parallelization on the beamlet level (Figure 2b, Part I), we used 128 threads per block and 256 blocks per grid; for the parallelization on the beamlet and voxel levels (Figure 2b,

Part II), we used 256 threads per block and the number of beamlets as the block size. Varying the CUDA kernel size had little impact on the execution time $T_{GPU}$. For example, for Case #3 listed in Table 1, the computational times were 0.17, 0.18 and 0.18 seconds when we changed threads number in Part II from 64, 128 to 256, respectively.

The computational time on the CPU ranged from 21 to 124 seconds. On the GPU, excluding data transfer time, the computational time is less than 0.5 seconds, which leads to a speedup factor around 400. The computational error for the GPU implementation is defined as $\varepsilon = \frac{1}{N}\sum_{i,j}\left|D_{ij}^{CPU} - D_{ij}^{GPU}\right|$, where $N$ is the total number of dose deposition coefficients. For all seven cases, the value of $\varepsilon$ is constantly around $10^{-6}$.

**Table 1**. Execution time and speedup factors for different beamlet sizes and voxel sizes. Here, $N'_{voxel}$ is the total number of voxels in all beamlets' VOIs.

| # | Voxel size (cm$^3$) | Beamlet size (cm$^2$) | $N_{voxel}$ ($\times 10^6$ voxels) | $N_{beamlet}$ (beamlets) | $N'_{voxel}$ ($\times 10^7$ voxels) | $T_{CPU}$ (sec) | $T_{GPU}$ (sec) | $T^*_{GPU}$ (sec) | Speedup $\frac{T_{CPU}}{T_{GPU}}$ | Speedup $\frac{T_{CPU}}{T^*_{GPU}}$ |
|---|---|---|---|---|---|---|---|---|---|---|
| 1 | $0.50^3$ | $0.20^2$ | 0.22 | 2500 | 8.64 | 21.22 | 0.06 | 0.11 | 373.04 | 192.32 |
| 2 | $0.37^3$ | $0.20^2$ | 0.51 | 2500 | 1.80 | 42.80 | 0.10 | 0.19 | 409.34 | 222.84 |
| 3 | $0.30^3$ | $0.20^2$ | 1.00 | 2500 | 3.23 | 78.27 | 0.18 | 0.34 | 419.80 | 230.22 |
| 4 | $0.25^3$ | $0.20^2$ | 1.73 | 2500 | 5.27 | 124.54 | 0.30 | 0.56 | 420.73 | 223.24 |
| 5 | $0.25^3$ | $0.25^2$ | 1.73 | 1600 | 4.97 | 120.14 | 0.29 | 0.53 | 414.85 | 225.91 |
| 6 | $0.25^3$ | $0.33^2$ | 1.73 | 900 | 4.70 | 112.78 | 0.27 | 0.46 | 416.40 | 244.13 |
| 7 | $0.25^3$ | $0.50^2$ | 1.73 | 400 | 4.39 | 100.77 | 0.24 | 0.43 | 417.10 | 232.61 |

When comparing $T_{GPU}$ to $T^*_{GPU}$, we observed that the data transfer time between CPU and GPU is comparable to the GPU computational time, which reduces the speedup factor from ~400 ($T_{GPU}$) to ~200 ($T^*_{GPU}$). This observation is not surprising, as it confirms the expensive nature of communication between CPU and GPU. Consequently, our strategy to minimize data transfer by storing data directly on the device appears to be correct.

*3.2 A clinical case*

The performance evaluation of our FSPB implementation on both GPU and CPU was also carried out for a prostate clinical case, where 9 co-planar 6MV beams were used. The beamlet size was $0.5^2$ cm$^2$, which yields 528~600 beamlets for each beam. The voxel size was 0.25 cm$^3$. For each beamlet, there were about $1.0$~$3.0 \times 10^4$ voxels involved in computation. Note that as in the water phantom case, there is essentially no difference between GPU and CPU calculations for this clinical case. The sequential CPU computation took about 4.8 minutes. For the parallel GPU implementation, it took 0.7 seconds for the computation only and 1.2 seconds including data transfer between CPU and GPU. We used the same CUDA kernel size as in the water phantom case.

The accuracy of the radiological depth is related to the distance (or the element length) between two neighboring interpolation points used in the lookup table. To test the accuracy of this lookup table, we sequentially decreased the distance between two neighboring points from the full length of a voxel to one half, one third, and one quarter. However, we did not observe any appreciable change in the final values of dose deposition coefficients. This may be due to the relatively small variation of the CT numbers in the prostate case. When applying our implementation to other tumor sites where CT image inhomogeneity is significant, we may need to use finer element length to build the radiological depth lookup table. For the results presented in this paper, we use half of the voxel dimension as the distance between two neighboring points. We would also like to point out that the GPU time is almost independent of the lookup table resolution; when the element length decreases by a factor of 4, the GPU time only increases by less than 1%.

**4. Discussion and Conclusions**

This paper presents the development of an FSPB-based GPU parallel computing framework for IMRT dose calculation. An analytical FSPB model utilizing three error functions as the FSPB kernel was used for illustration purposes. Any other FSPB models can be easily implemented in our GPU framework by simply replacing a few lines in the CUDA code with an alternative FSPB kernel. FSPB models are particularly suitable for computing dose deposition coefficients for IMRT plan optimization. For most clinical cases, their accuracy is sufficient. However, more advanced dose calculation models may be needed when there exists large inhomogeneities. For this reason, we are also developing a GPU-based parallel platform for Monte Carlo dose calculation.

In this study, the computational speed and numerical accuracy of the GPU code were tested both with a water phantom and a clinical case. The computational accuracy evaluation has shown that the error is well controlled on the level of single floating-point precision of $10^{-6}$, which is negligible in real clinical applications.

The achieved speedup gain from using GPUs is affected by a number of factors. First, the speedup depends primarily on hardware configuration. We tested our CUDA code on four different types of NVIDIA GPUs: GeForce 9500 GT, GTX 285, Tesla C1060, and Tesla S1070. Among them, GeForce 9500 GT has only 4 multiprocessors. Both GTX 285 and C1060 have 30 multiprocessors. S1070 has 4 GPUs, each with 30 multiprocessors. The clock rates of these four cards are very similar varying from 1.3 GHz to 1.48 GHz. When performing the water phantom example Case #1 (as shown in Table 1) on a single GPU, we found that $T_{GPU}$ values for GTX 285, C1060 and S1070 are nearly identical, which are, however, over 7 times faster than that for GeForce 9500 GT. This result confirms that the number of multiprocessors plays a fundamental role in deciding the computation time when the clock rate is fixed.

Another major factor affecting computational speed is the size of GPU memory. GeForce 9500 GT has 512 MB global memory, GTX 285 has 1GB memory, and both C1060 and S1070 have 4GB memory per GPU. On C1060 and S1070, for all testing scenarios, the computation can be executed in one pass, requiring only two data transfers between the CPU and the GPU, once each at the beginning and end of GPU computation. On GTX 285 and GeForce 9500 GT, for Cases #4, #5, #6 and #7 in Table 1, the computation cannot be done all at once for 5 beams. After finishing the calculation of dose deposition coefficients for 2 or 3 beams on the GPU, we had to copy the results back to the CPU in order to free the GPU memory for computation over the remaining beams. As mentioned above, the bandwidth between the GPU and the CPU is much lower than that within the GPU, so the additional CPU-GPU data transfer will significantly erode efficiency gains. Choosing the proper GPU for online adaptive therapy is based on a trade-off between cost and performance. Based on the currently available GPUs, we recommend the use of NVIDIA Tesla C1060 cards, which cost around $1,500 and can be readily inserted into a workstation PC.

The speedup factor also depends on the implementation of the FSPB model on CPU. Our CPU implementation is comparable to that of Jelen *et al* (2005). The computation time for the CPU code can certainly be reduced by optimizing the code using various well established tricks. We did not focus on this issue since for the purpose of this project, what matters the most is the GPU computation time, not the CPU time nor the speedup factor.

The dose deposition coefficient calculation is one of the most significant components in IMRT plan optimization and the main constraint for real-time online adaptive radiotherapy. In this work, the dose deposition coefficient calculation was performed independently on individual beamlets and voxels. The parallel nature of dose calculation warrants significant acceleration by massively parallel computing on GPU. Our GPU implementation of the FSPB model allows the computation of a 9-field prostate IMRT plan *within one second*. The unprecedented speedup factors achieved in this work clearly demonstrates a feasible and promising solution to the realization of real-time online re-planning for adaptive radiotherapy.

**Acknowledgements**

This work is supported in part by the University of California Lab Fees Research Program. We would like to thank NVIDIA for providing GPU cards and Dr. Lai Qi for his help with figures.